\documentstyle[12pt]{article}
\newcommand{\be}{\begin{equation}}
\newcommand{\ee}{\end{equation}}
\newcommand{\ba}{\begin{array}}
\newcommand{\ea}{\end{array}}

\renewcommand{\vec}[1]{{\bf #1}}

\begin{document}
%\twocolumn

\title{DMRG study of ferromagnetism in a one-dimensional Hubbard model}
\author {\rm S.\ Daul$^1$ and R.M.\ Noack$^2$ \\ 
$^1$Institut de physique th\'{e}orique, Universit\'{e} de Fribourg, \\
   CH-1700 Fribourg, Switzerland \\
   $^2$Institut f\"ur Theoretische Physik, Universit\"{a}t
   W\"{u}rzburg,
   Am Hubland, \\
   97074 W\"{u}rzburg, Germany  \\
}
\maketitle

\begin{abstract}

The one dimensional Hubbard model with nearest and (negative) next-nearest 
neighbour hopping has been studied with the density-matrix renormalization group (DMRG)
method. A large region of ferromagnetism has been found for finite density
and finite on-site interaction. 


\end{abstract}

\section{Introduction}

The Hubbard model was originally introduced to describe
correlation effects in transition metals, such as, for example, the band
ferromagnetism of Fe, Co and Ni. 
In mean field theory, one finds ferromagnetism
whenever the Stoner criterion is satisfied, 
which, in the Hubbard model, leads to substantial regions of 
ferromagnetism in the phase diagram.
When correlation effects are taken into account, however, 
a fully polarized ground state is quite difficult to find. 
For dimension $d=1$, a theorem proven by Lieb and Mattis requires the
ground state of the Hubbard model with near-neighbor hopping to be a
singlet, completely excluding ferromagnetism \cite{lieb_mattis}.
For $d \geq 2$, ferromagnetism is possible via the Nagaoka
mechanism at very large interaction and near half--filling, but has
not been found elsewhere in the phase diagram.
Here we concentrate on the 1D case which can be extended to
circumvent the Lieb--Mattis theorem so that ferromagnetism is
allowed. One way to do this is to add orbital degeneracy to the model 
in order to mimic the Hund's rule ferromagnetism found in atoms;
another way is to add interaction terms such as
nearest-neighbour Coulomb terms \cite{vollhardt}. 
Finally, one can change the band structure by introducing 
a ``flat band'' \cite{tasaki} or, as is done here, by adding 
longer range hopping terms.

%------------------------------------------------------------------------------
\section{Model and method}

We will study the 1D Hubbard model with nearest and next-nearest neighbour
hopping: 
\be
   H = - \sum_{i=1,\sigma=\uparrow,\downarrow}^L (t_1 c^{\dag}_{i\sigma}c_{i+1\sigma} +
      t_2 c^{\dag}_{i\sigma}c_{i+2\sigma} + \mbox{h.c.}   )
      + U \sum_{i=1}^L n_{i\uparrow}n_{i\downarrow}.
\ee
Here $c^{\dag}_{i\sigma}$ creates an electron of spin $\sigma$ on site
$i$, $n_{i\sigma} =c^{\dag}_{i\sigma} c_{i\sigma}$, 
$L$ is the system size, and $U$ is the on-site Coulomb
interaction.
In the following we express all energies in units of $t_1=1$ 
and only consider negative values of $t_2$. 
Because a definite order of the particles is no longer enforced when
$t_2 \ne 0$, the Lieb--Mattis theorem does not apply 
and, indeed, ferromagnetism has analytically been shown to exist  at
$U=\infty$ in three different limits:
For one hole in a half-filled band Nagaoka ferromagnetism has been found
\cite{mattis_pena}; for $|t_2| \rightarrow 0$, it has been shown 
\cite{ueda} that the model is ferromagnetic for all densities;
and for $|t_2|>0.25$, where the band structure has two minima, 
M\"{u}ller-Hartmann \cite{mh} has shown that the low density limit is
ferromagnetic.
These three limits are indicated in the schematic phase diagram shown in
Fig. 1. 
In addition, for $|t_2| \rightarrow \infty$ the model can be mapped onto
two decoupled Hubbard chains, which cannot be ferromagnetic due to the
Lieb--Mattis theorem. 
The aim of this work is to determine the extent of these domains of
ferromagnetism in $n$ and $U$, and whether or not these regimes are
connected.

Previous work using Lanczos exact diagonalization and
variational techniques with various trial functions \cite{pieri,daul}
has shown that the ferromagnetic domain is large.
However, the exact diagonalization calculations in this work showed
large finite size effects associated with closed shell effects in
momentum space.
Here we use the density matrix renormalization group (DMRG) method
developed by White \cite{white} and now widely used,
to obtain results for much larger system sizes using open boundary
conditions.

%------------------------------------------------------------------------------
\section{Results}

Since the ground state energy obtained in the DMRG, $E_D$, is
variational, it provides an upper bound for the exact ground state energy.
We can analyse the stablity of the fully polarized state by comparing
its energy, $E_F$, which is exactly known since the state
has no double occupancy, with  $E_D$.
For a given value of $t_2$ and density $n$ we can find a critical
value $U_c$ below which the ferromagetic state is unstable. 
Since  $E_D$ can be found with high precision for this
model, $U_c$ can be determined  accurately.
To check this we have calculated the total spin $S$ in the DMRG ground
state by evaluating
\be
   \langle \vec{S}^2 \rangle =
    \sum_{i,j} \langle \vec{S}_i \vec{S}_j \rangle.
\ee
We find that the value of $S$ goes smoothly from 0 to the fully polarized value
$S_{\mbox{max}}$ at the $U_c$ found by comparing the energies.
It is difficult to determine 
whether the change in $S$ is continous or not.
The near--degeneracy of states with different $S$ at the transition
leads to mixing of states  in the
diagonalization step of the procedure, so that $\langle \vec{S}^2 \rangle$
no longer takes on definite discrete values.
Using these two tools we have calculated the $U=\infty$ phase
diagram shown in Fig. 2.
The full dots are points at which the ground state is
fully polarized (i.e. $E_D > E_F$ and $S \approx S{\mbox{max}}$ ); 
the empty squares are points where the system is not magnetic (i.e. 
$E_D < E_F$ and  $S \approx 0$); and 
the dashes are points at which the ferromagnetism is not fully
saturated (i.e. $E_D > E_F$ and $S \neq S{\mbox{max}}$). 
We can see that the previous analytical limits are well reproduced in this 
phase diagram and that the regions associated with the three
mechanisms are connected.

We next examine the behavior at finite $U$.
In Fig. 3, we choose three representative values 
of $t_2$ $(-0.1, -0.8, -2.0)$, and show $U_c$ as a function of
the density $n$. 
At $t_2=-0.1$, the band has only one minimum, at $t_2=-2.0$ 
the ferromagnetic region does not occur for all densities at
$U=\infty$, and $t_2=-0.8$ is intermediate between the two regions.
We see that for the two cases in which there are two
minima $(|t_2| > 0.25 )$ in the band structure, there is a local minimum in $U_c$ at a
given density $n_c$.
This density corresponds to the Fermi level being just at the top of
the barrier between the two band minima, which leads to a high density
of states, favourable for ferromagnetism.

%------------------------------------------------------------------------------
\section{Conclusion}

We have found a ferromagnetic ground state in a large parameter
regime in the 1D Hubbard model with next-nearest neighbor hopping,
showing that ferromagnetic regimes found in particular limits at
$U=\infty$ extend to finite $U$ and  density.
We also find that these different ferromagnetic regimes are connected.
One interesting unresolved issue is whether the transition from a
paramagnetic to a ferromagnetic state is continuous.
This issue can be clarified in future work
by looking at appropriate correlation functions in more detail.
The present results confirm  that the ``critical density'' $n_c$ introduced in
previous work \cite{pieri} does not represent a true phase boundary \cite{daul},
but merely a density where the $U$
value needed to stabilize ferromagnetism becomes small.

%------------------------------------------------------------------------------
\section{Acknowledgments}

It is a pleasure to thank D. Baeriswyl and M. Dzierzawa for useful discussions.
S. D. was supported by the Swiss National
Foundation through grant No. 20-40672.94.

%------------------------------------------------------------------------------

%------------------------------------------------------------------------------

{\bf Figure captions}

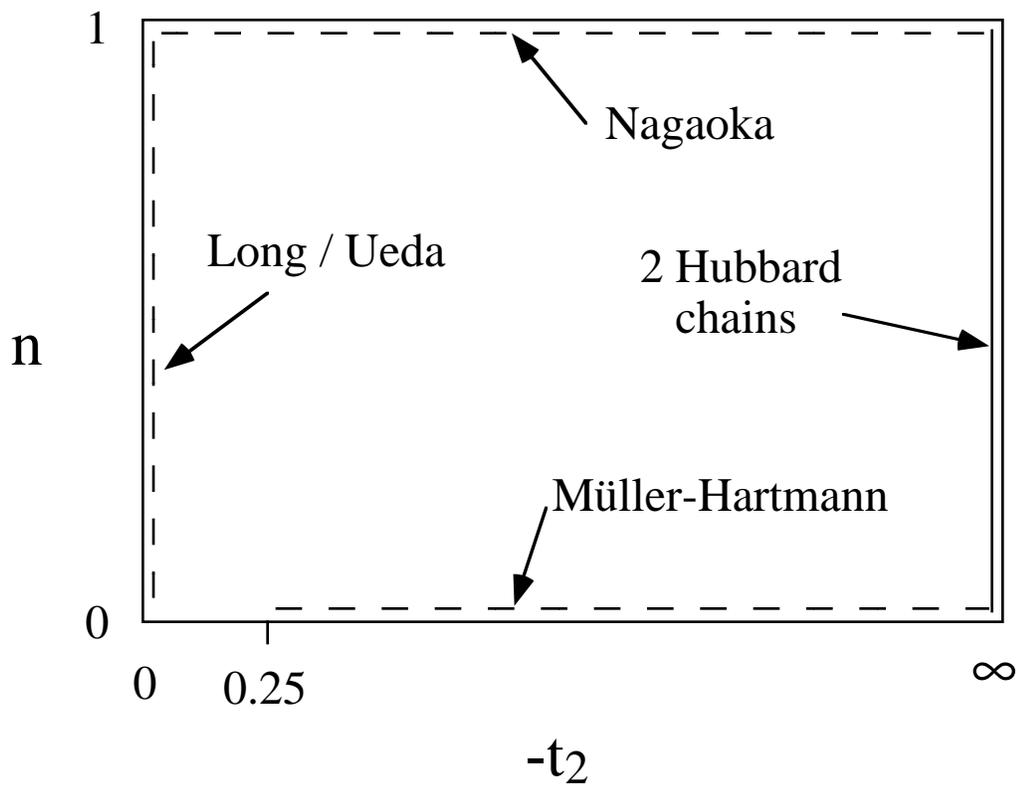
{\bf Fig. 1}

Schematic $U = \infty$ ground state phase diagram in $n-t_2$ plane. 

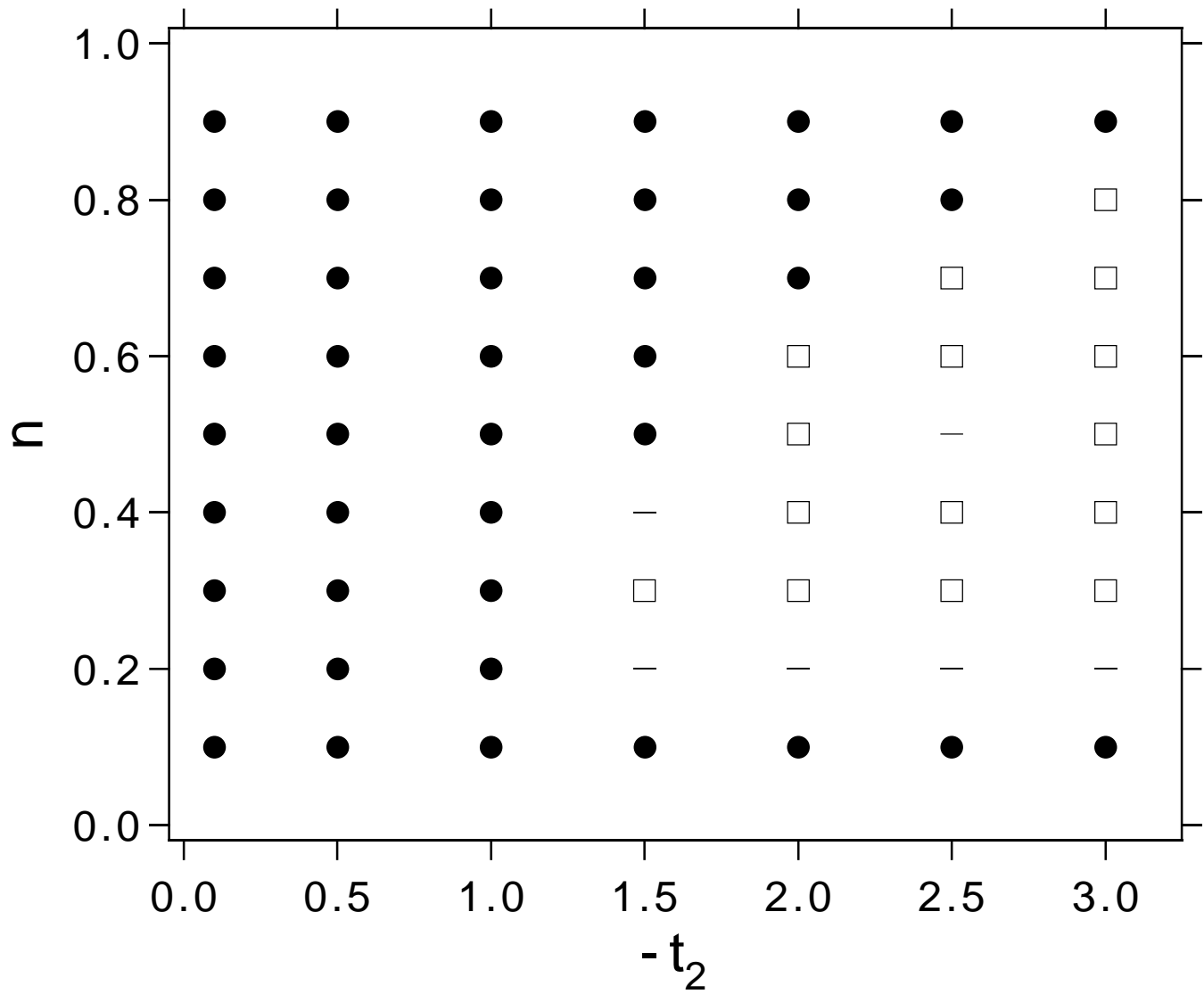
{\bf Fig. 2}

The full dots show a ferromagetic groundstate, the open squares are for
paramagnetic ground state, and the dashes are for points where doubts
remain. The energy and $S$ where calculated on a system of size
$L=30$ at $U=10^6$.

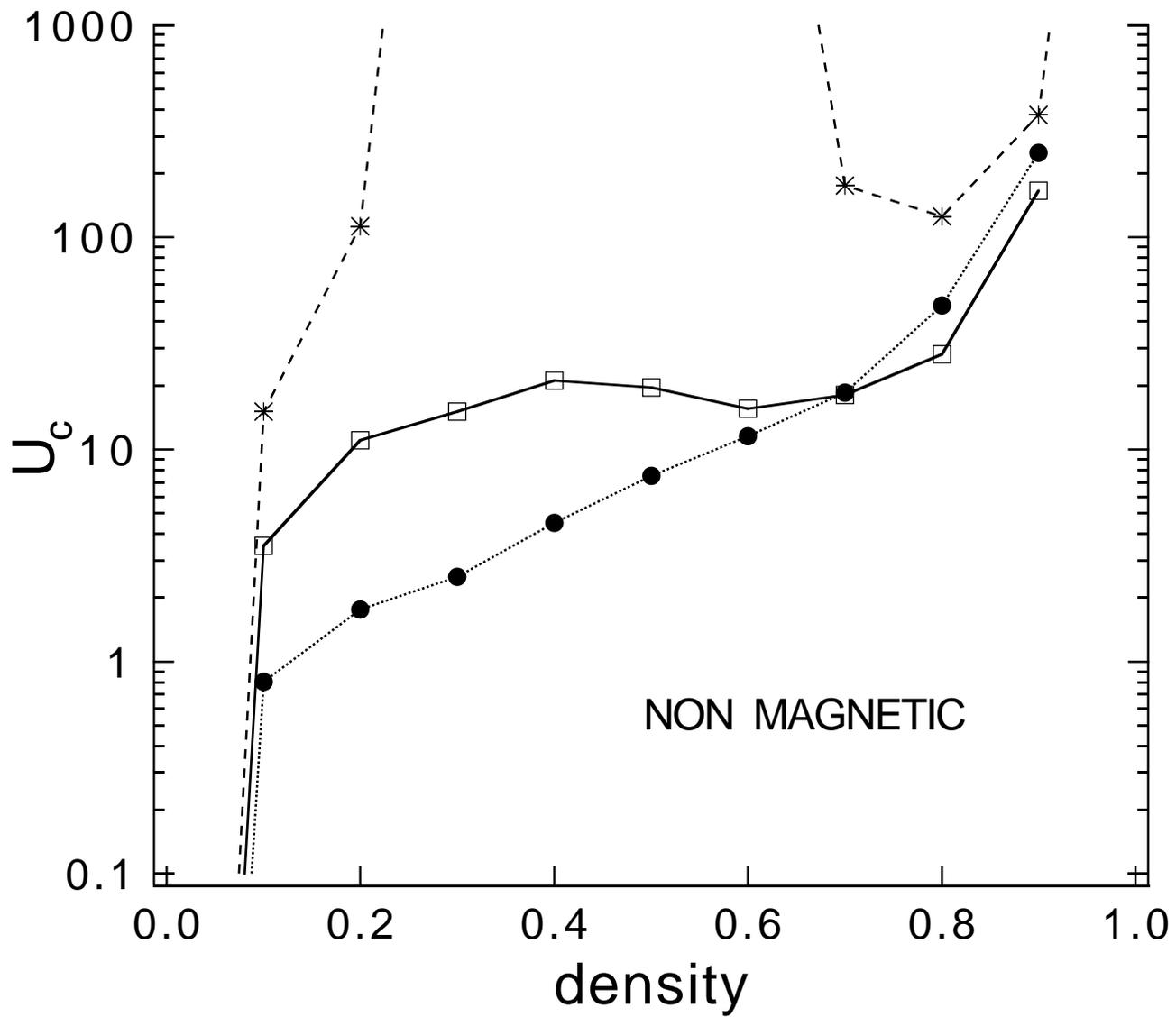
{\bf Fig. 3}

Critical value of $U$ above which ferromagnetism is found for three different
values of $t_2$, namely  $t_2=-0.2$ (full dots), $t_2=-0.8$ (open squares)
and $t_2=-2$ (stars). The lattice size is $L=50$.

\end{document}